\documentclass[12pt]{article}
\usepackage{amsmath,amssymb,amsfonts}
\usepackage[paper=letterpaper,margin=1.0in]{geometry}
\usepackage{graphicx}

\newcommand{\be}{\begin{equation}}
\newcommand{\ee}{\end{equation}}
\newcommand{\bea}{\begin{eqnarray}}
\newcommand{\eea}{\end{eqnarray}}
\newcommand{\ba}{\begin{array}}
\newcommand{\ea}{\end{array}}
\newcommand{\ben}{\begin{enumerate}}
\newcommand{\een}{\end{enumerate}}
\newcommand{\bi}{\begin{itemize}}
\newcommand{\ei}{\end{itemize}}
\newcommand{\bc}{\begin{center}}
\newcommand{\ec}{\end{center}}
\newcommand{\bfig}{\begin{figure}}
\newcommand{\efig}{\end{figure}}
\newcommand{\bq}{\begin{quotation}}
\newcommand{\eq}{\end{quotation}}
\newcommand{\bt}{\begin{table}}
\newcommand{\et}{\end{table}}
\newcommand{\btab}{\begin{tabular}}
\newcommand{\etab}{\end{tabular}}
\newcommand{\bs}{\begin{slide}}
\newcommand{\es}{\end{slide}}

\newcommand{\AdS}[1]{{\rm AdS}_{#1}}
\newcommand{\pa}{\partial}

\begin{document}

{\footnotesize
${}$
}

\bc

\vskip 1.0cm
\centerline{\Large \bf Quantum Gravity and Turbulence}
\vskip 0.5cm
\vskip 1.0cm

\renewcommand{\thefootnote}{\fnsymbol{footnote}}

\centerline{{\bf
Vishnu Jejjala${}^{1}$\footnote{\tt v.jejjala@qmul.ac.uk},
Djordje Minic${}^{2}$\footnote{\tt dminic@vt.edu},
Y.\ Jack Ng${}^{3}$\footnote{\tt yjng@physics.unc.edu}, and
Chia-Hsiung Tze${}^{2}$\footnote{\tt kahong@vt.edu}}}

\vskip 0.5cm

{\it
${}^1$Centre for Research in String Theory \\ Department of Physics, Queen Mary, University of London \\ Mile End Road, London E1 4NS, U.K. \\
${}$ \\
${}^2$Institute for Particle, Nuclear and Astronomical Sciences \\ Department of Physics, Virginia Tech \\ Blacksburg, VA 24061, U.S.A. \\
${}$ \\
${}^3$Institute of Field Physics \\ Department of Physics and Astronomy, University of North Carolina \\ Chapel Hill, NC 27599, U.S.A.
}

\ec

\vskip 1.0cm

\begin{abstract}
We apply recent advances in quantum gravity to the problem of turbulence.
Adopting the AdS/CFT approach we propose a string theory of turbulence that explains the Kolmogorov scaling in $3+1$ dimensions and the Kraichnan and Kolmogorov scalings in $2+1$ dimensions.
In the gravitational context, turbulence is intimately related to the properties of spacetime, or quantum, foam.
\end{abstract}

\vspace{1cm}

\begin{center}

This Essay received Honorable Mention in the 2010 Gravity Research Foundation Essay Contest 

\end{center}

\renewcommand{\thefootnote}{\arabic{footnote}}

\newpage

\section{Classical Diffeomorphisms and Metric Structures}

John Wheeler was among the first to realize the connections between quantum gravity and the ubiquitous phenomenon of turbulence.
Due to quantum fluctuations, spacetime, when probed at very small scales, will appear very complicated --- something akin in complexity to a chaotic turbulent froth, which Wheeler dubbed a {\em quantum foam}, also known as the {\em spacetime foam}.
In this essay, we explore the deep similarities and cross fertilization between quantum gravity and turbulence, culminating in a string theory of turbulence~\cite{previous}.

The basic dynamical equation for turbulent flow is the non-linear Navier--Stokes equation
$\rho ( \pa_t v_i + v_j\, \pa_j v_i ) = -\pa_i p + \nu\, \pa_j^2 v_i$,
with the incompressibility condition $\pa_i v_i = 0$, where the velocity field of the flow is $v_i$, $p$ is pressure, and $\rho$ is the fluid density~\cite{review}.
This can be understood as Newton's second law for a fluid and probably embodies the physics of any fluid flow.
We are interested in the most complicated flows of all, namely {\em fully developed turbulence}, or turbulence in the limit of infinite Reynolds number $R$.
As $R = L v/\nu$, where $L$ is a characteristic scale and $\nu$ is the kinematic viscosity, we look at the limit of vanishing viscosity.
In this regime, all the various possible symmetries are restored in a statistical sense, calling for a probabilistic description of what is in essence a deterministic system.
Therefore, in computing correlators of the velocity field, we are led to make use of the statistical and Euclidean quantum field theoretic descriptions of turbulence.

The connection between quantum gravity and turbulence should not come as a surprise when one recalls the role of the (volume preserving) diffeomorphism symmetry in classical (unimodular) gravity and the volume preserving diffeomorphisms of classical fluid dynamics.
Furthermore, in the case of irrotational fluids in three spatial dimensions, the equation for the fluctuations of the velocity potential can be written in a geometric form~\cite{unruh} with a harmonic Laplace--Beltrami equation:
$\frac{1}{\sqrt{-g}} \partial_a( \sqrt{-g} g^{ab} \partial_b \varphi) = 0$.
Apart from a conformal factor, the acoustic metric has the canonical ADM form~\cite{unruh,abh}:
${\rm d}s^2 = \frac{\rho_0}{c} [ c^2 dt^2 - \delta_{ij}(dx^i - v^i dt)(dx^j - v^j dt)]$,
where $c$ is the sound velocity and $v^i$ are the components of the fluid's velocity vector.
We observe that in this expression for the metric, the velocity of the fluid $v^i$ plays the role of the shift vector $N^i$, which is the Lagrange multiplier for the spatial diffeomorphism constraint (the momentum constraint) in the canonical Dirac/ADM treatment of Einstein gravity:
${\rm d}s^2 = N^2 dt^2 - h_{ij} (dx^i + N^i dt) (dx^j + N^j dt)$.
In the fluid dynamics context, $N^i \rightarrow v^i$, and a fluctuation of $v^i$ would imply a quantum fluctuation of the shift vector.
Herein lies a natural mapping between scalings in turbulence and spacetime foam, allowing us to uncover the universal properties of the latter.

\section{Kolmogorov Scaling and Holographic Quantum Foam}

In fully developed turbulence in three spatial dimensions, there is the remarkable {\em Kolmogorov scaling}, which specifies the behavior of $n$-point correlation functions of the fluid velocity.
Kolmogorov scaling~\cite{kol} follows from the assumption of constant energy flux, $\frac{v^2}{t} \sim \varepsilon$, where the single length scale $\ell$ is given as $\ell \sim v\cdot t$.
This implies that
$$
v \sim (\varepsilon\, \ell)^{1/3}~,
$$
consistent with the experimentally observed two-point function $\langle v^i(\ell) v^j(0)\rangle \sim (\varepsilon\, \ell)^{2/3} \delta^{ij}$, which yields, via a one-dimensional Fourier transform, the energy scaling $E(k) \sim k^{-5/3}$ law.

On the other hand, at microscopic scales our world is known to obey quantum mechanics which is characterized by an indeterminacy giving rise to fluctuations in measurements.
If spacetime, like all matter and energy, undergoes quantum fluctuations, there will be an intrinsic limitation to the accuracy with which one can measure a distance $\ell$, for that distance fluctuates by $\delta \ell$.
Applying quantum mechanics and black hole physics (from general relativity) to a gedanken experiment to measure a distance $\ell$, one can show that $\ell$ fluctuates by an amount $\sim \ell^{1/3} \ell_P^{2/3}$, where $\ell_P$ is the Planck length, the characteristic length scale in quantum gravity~\cite{ng}.
The corresponding quantum foam model has become known as the holographic model since it can be shown to be consistent with the holographic principle.
If one defines a velocity as $v \sim \frac{\delta \ell}{t_c}$, where the natural characteristic time scale is $t_c \sim \frac{\ell_P}{c}$, then it follows that
$$
v \sim c \big(\frac{\ell}{\ell_P}\big)^{1/3}~.
$$
It is now obvious that we have obtained a Kolmogorov-like scaling \cite{previous} in turbulence,
{\em i.e.}, the velocity scales as $v \sim \ell^{1/3}$.
Since the velocities play the role of the shifts, they describe how the metric fluctuates at the Planck scale.
The implication is that at short distances, spacetime is a chaotic and stochastic fluid in a turbulent regime with the Kolmogorov length $\ell$~\cite{previous}.
Thus, in $3+1$ dimensions, holography and turbulence appear to be in harmony.

\section{Wilson Loop and Kolmogorov/Kraichnan Scalings}

For gauge theories like QCD, gauge invariance can be naturally exhibited by the use of loop variables.
Likewise, for turbulence, symmetries such as area and volume preserving {\em quantum diffeomorphisms} can be simply incorporated by the use of the Wilson loops (to be later manifested in the form of exponents of a power of area or volume in the expectation value of the turbulent loop functional).
Let us follow Migdal~\cite{migdal} in rewriting the Navier--Stokes equation as an effective Schr\"odinger equation involving the Hopf loop functional,
{\em i.e.}, the ``turbulent Wilson loop''
$$
W(C) \sim \exp\left(-\frac{1}{\nu}\int_C dx_i\ v_i \right) ~,
$$
with the viscosity $\nu$ playing the role of $\hbar$.
In the WKB approximation with $\nu \rightarrow 0$, there exists a self-consistent scaling law.
For the case of Kolmogorov scaling, inserting this self-consistent ansatz into the turbulent Wilson loop, we find
$$
W_{\rm Kol} \sim \exp\left(-\frac{\alpha}{\nu}\, \varepsilon^{1/3}\, A^{2/3}\right)
$$
since $A \sim \ell^2$.
(The undetermined real prefactor $\alpha$ absorbs dimensionful constants.)

In $2+1$ dimensions, in addition to energy there is a second conserved quantity, the enstrophy $\Omega = \int d^2x\ \omega^2$, where $\omega$ is the vorticity vector $\vec{\omega} \equiv \nabla \times \vec{v}$.
According to Kraichnan~\cite{kr}, the constant flux of enstrophy gives $\frac{\omega^2}{t} \sim {\rm constant}$ and implies that the statistical velocity field scales as $v \sim \frac{\ell}{t_0}$, where $t_0$ is the characteristic constant.
This leads to the $k^{-3}$ scaling of the energy in momentum space.
In $2+1$ dimensions we have both the energy (Kolmogorov) and the enstrophy (Kraichnan) cascades.
Given the Kraichnan scaling, the turbulent Wilson loop goes as
$$
W_{\rm Kr} \sim \exp\left(-\frac{\alpha}{\nu\, t_0}\, A\right) ~.
$$

Two remarks are in order.
Firstly, we note that both the Kolmogorov and the Kraichnan scalings are related to the constancy of the three-point function, $\langle v_a v_b \partial_a v_b \rangle \sim {\rm constant}$ and $\langle v_a \omega_b \partial_a \omega_b \rangle \sim {\rm constant}$, respectively.
This allows for the interpretation of both scalings in terms of a quantum field theoretic anomaly~\cite{polyakov,gawedzki}.
Secondly, a naive application of holography in $2+1$ dimensions yields a $k^{-2}$ energy scaling~\cite{previous}, different from that given by Kraichnan scaling.
The resolution of this conflict is found in a new picture of turbulence provided by string theory to which we now turn.

\section{String Theory and Turbulence}

The above scaling of the Migdal--Wilson loop in the Kraichnan regime gives the same area law as in the case of confining QCD.
This result readily hints at a connection between string theory and turbulence, at an underlying effective Nambu--Goto action ($S_{\rm NG}$) of string theory.
Sure enough, a standard calculation in AdS/CFT~\cite{adscft,maldacena} yields the expectation value of the turbulent Wilson loop in the Kraichnan scaling regime in $2+1$ dimensions;
it satisfies an area law:
$$
\langle W(A) \rangle = \exp(-S_{NG}) = \exp(-f\, A) ~.
$$
(The prefactor $f$ absorbs the dimensionful constants.)
Hence the boundary turbulence in the Kraichnan regime is given by string theory with the Nambu--Goto action in the bulk of $\AdS{4}$.


We now show that, in our string theory of turbulence, both the Kraichnan and the Kolmogorov scalings can be understood from a unified point of view~\cite{previous}.
Starting with the fluid vortex dynamics, for a single big vortex we have an effective action given by the Nambu--Goto action.
Now, if this vortex is turned into two, and then four, etc., at the end of the cascade we will have a large number of small vortices.
This means that the area spanned by the vortex is now made of many little vortex areas.
The worldsheet has, from a coarse grained point of view, become effectively a worldvolume.
In terms of the original Nambu--Goto action for one big vortex we have
$$
\exp(-f\, A) \sim \exp(-f\, V^{2/3}) ~.
$$
This result is what one expects if turbulence can be formulated as an effective string theory.
The point is that at {\em strong coupling} the turbulent string would turn into a membrane in the same way the usual fundamental string turns into a membrane in M-theory~\cite{dhis}.
From the membrane point of view, the $\exp(-f\, V^{2/3})$ result becomes effectively $ \exp(-f\, A)$, {\em i.e.}, we have the area law for Kraichnan scaling.
The only fact we need to use in this dynamical picture is volume preserving diffeomorphisms as the big vortex decays into many small vortices.
Furthermore, if one allows for a {\em membrane/string transition}, by lowering the string coupling we go from $\exp(-f\, V^{2/3})$ to $\exp(-f\, A^{2/3})$, and this gives the Kolmogorov scaling.
This shows that the $2/3$ exponent is common to both the Kolmogorov and Kraichnan scalings, and the observed Kolmogorov scaling emerges as a consequence of string theory.

The $2/3$ exponents also allow us to view the Kolmogorov/Kraichnan scalings as deformations of the area/volume laws.
Comparing $W_{\rm Kol}$ to the usual area law of the Yang--Mills theory, we can regard the Kolmogorov result as a turbulent {\em deformation} of the area law, so that
$$
\exp(-f\, A)\ \Longrightarrow\ \exp(-f\, A^{2/3}) ~.
$$
For the three-dimensional Yang--Mills theory, which is not conformal as the coupling is dimensionful, we would have the same deformation.
But according to the AdS/CFT correspondence, in three dimensions we can have another theory --- the theory of interacting membranes.
For the membrane theory, dual to M-theory on $\AdS{4}$, we should have a volume law associated with Wilson surface operators.
Then we can view the Kraichnan scaling as a turbulent {\em deformation} of the volume law
$$
\exp(-f\, V)\ \Longrightarrow\ \exp(-f\, V^{2/3}) ~.
$$

The Kraichnan and Kolmogorov scalings in $2+1$ dimensions have a natural $3+1$ dimensional counterpart.
In $3+1$ dimensions, we only have the Yang--Mills theory, the usual four-dimensional gauge theory dual to some string theory on an $\AdS{5}$ space with a cutoff.
In this case we only have the deformation of the area law and thus the Kolmogorov scaling.
This $\exp(-A^{2/3})$ law would just be the {\em dimensional lift} of the same law in $2+1$ dimensions.
As we do not have a membrane theory in $3+1$ dimensions, there would be no counterpart to Kraichnan scaling.
Conversely, the three-dimensional case can be viewed as a {\em dimensional reduction} of the four-dimensional case.
When we reduce a turbulently deformed area law from four dimensions to three dimensions, we get either the same result or its volume analogue, where the strings from three-dimensional Yang--Mills thicken into membranes~\cite{tassos}.
On top of this we have a natural inverse RG provided by the holographic RG relation between the boundary~\cite{rg} (where the turbulent string is) and the bulk (where the fundamental string is).


Finally, this picture implies the bulk string theory/boundary turbulence dictionary for the generating functional of velocity correlators as in the AdS/CFT correspondence.
The generating functional of all turbulent correlators of a fluid in the Kraichnan regime in $2+1$ dimensions is given as a bulk string partition function in the semiclassical regime with the viscosity $\nu$ playing the role of an expansion parameter ({\em i.e.}, some sort of effective $\hbar$).
Kraichnan turbulence in $2+1$ dimensions is dual to string theory in $\AdS{4}$.

\section{Conclusions and Open Problems}

To recapitulate, we have shown that there are deep connections between quantum gravity and turbulence.
In $3+1$ dimensions, the Kolmogorov scaling in turbulence is intimately related to the properties of spacetime fluctuations.
But at first sight, there appears to be some friction between holography and Kraichnan scaling in $2+1$ dimensions.
To resolve that conflict we have proposed a string theory of turbulence in which, reinterpreted from the AdS point of view, the holography is not on the boundary but in the bulk.
This proposal explains the Kolmogorov scaling in $3+1$ dimensions and the relationship between the Kraichnan and Kolmogorov scalings in $2+1$ dimensions.
We speculate that the universal $2/3$ exponent is an indication that one is working in the spacetime foam regime (from a boundary point of view)~\cite{previous}.
Not only is string theory useful in formulating a theory of turbulence, but the physics of turbulence sheds light, as is done here through scaling laws, on the spacetime foam phase of strong quantum gravity.
We have uncovered a curious synergy between the two seemingly disparate fields of quantum gravity and turbulence.
In so doing, we have also made a foray into the terra incognita of non-equilibrium phenomena in quantum gauge and gravity theories.

Our new picture of turbulence opens up many new avenues for future investigation.
Among them is the problem of the matrix model reduction of our AdS/CFT dual and the connection to the matrix model for turbulence of~\cite{raj1}.
Another is the problem of loop equations as the anomaly equations in loop space~\cite{raj2}.
As general anomalous scaling laws appear in magnetohydrodynamics~\cite{zeldo} and other dynamical systems~\cite{kardar}, these physical settings provide fertile ground for study.
Another problem involves some subtleties concerning higher point correlators in fractal scalings due to intermittency~\cite{zeldo}.
Yet one more direction is the relation between Chern--Simons theories, the fractional quantum Hall effect, and fluids~\cite{lenny}.
Note that the boundary turbulent theory is a CFT (and thus similar to~\cite{polyakov}), but its correlator is given in terms of a bulk string theory.
A natural question is to consider the relation (if any) with the conformal fluid explored in~\cite{min}.
A more general lesson of our work may apply to the AdS/condensed matter correspondence.
One of the major puzzles in the application of AdS/CFT to condensed matter physics~\cite{hh} is why this should even work.
Our approach to turbulence employing the Wilson--Migdal loop may provide a clue.
The methods described in this essay map a new path toward a systematic understanding of turbulence.

\vskip 0.5cm

\noindent
{\bf Acknowledgments:}
VJ is supported by STFC.
DM is supported in part by the U.S.\ Department of Energy under contract DE-FG05-92ER40677.
YJN is supported in part by the U.S. Department of Energy under contract DE-FG02-06ER41418.

\end{document}